\documentclass[acmlarge]{acmart}

\AtBeginDocument{%
  \providecommand\BibTeX{{%
    \normalfont B\kern-0.5em{\scshape i\kern-0.25em b}\kern-0.8em\TeX}}}

\setcopyright{acmcopyright}
\copyrightyear{2020}
\acmYear{2020}
\acmDOI{10.1145/1122445.1122456}

\acmJournal{POMACS}
\acmVolume{37}
\acmNumber{4}
\acmArticle{111}
\acmMonth{5}

\usepackage[figuresleft]{rotating}
\usepackage{pdflscape}
\usepackage{tabulary}                                                           
\usepackage{amsmath}
\usepackage{booktabs}
\usepackage{amsmath,amsfonts}
\usepackage{algorithmic}
\usepackage{graphicx}
\usepackage{subcaption} 
\usepackage{tabularx}
\usepackage{textcomp}
\usepackage{longtable}
\usepackage{amsmath}
\usepackage{xcolor,colortbl}
\usepackage{xcolor}
\usepackage{array}
\usepackage{hyperref}
\def\BibTeX{{\rm B\kern-.05em{\sc i\kern-.025em b}\kern-.08em
   T\kern-.1667em\lower.7ex\hbox{E}\kern-.125emX}}
\newcolumntype{b}{X}
\newcolumntype{s}{>{\hsize=.5\hsize}X}
\begin{document}

\title{The Homophily Principle in Social Network Analysis}

\author{Kazi Zainab Khanam}
\email{kkhanam@lakeheadu.ca}
\affiliation{%
  \institution{Lakehead University}
  \city{Thunder Bay}
  \country{Canada}
  }
  
  \author{Gautam Srivastava}
\email{srivastavag@brandonu.ca}
\affiliation{%
	\department{Research Centre for Interneural Computing}
  \institution{China Medical University}
  \city{Taichung}
  \country{Taiwan}}
\affiliation{%
	\department{Department of Mathematics and Computer Science}
  \institution{Brandon University}
  \city{Brandon}
  \country{Canada}}

  \author{Vijay Mago*}
\email{vmago@lakeheadu.ca}
\affiliation{%
  \institution{Lakehead University}
  \city{Thunder Bay}
  \country{Canada}
  }

\renewcommand{\shortauthors}{Khanam, et al.}

\begin{abstract}
In recent years, social media has become a ubiquitous and integral part of social networking. One of the major attentions made by social researchers is the tendency of like-minded people to interact with one another in social groups, a concept which is known as Homophily. The study of homophily can provide eminent insights into the flow of information and behaviors within a society and this has been extremely useful in analyzing the formations of online communities. In this paper, we review and survey the effect of homophily in social networks and summarize the state of art methods that has been proposed in the past years to identify and measure the effect of homophily in multiple types of social networks and we conclude with a critical discussion of open challenges and directions for future research. 

\end{abstract}




\maketitle

\section{Introduction}

Homophily is a well-established phenomenon that has been observed to occur frequently in social networks, where users with similar contexts have a nature of connecting with one another constantly, and this principle is also a meticulously thought-out field in the domain of social sciences \cite{ma2015latent,pan2019twitter,zhu2015social,halberstam2016homophily,han2018analyze}. Homophily is a social concept where people’s personal networks tend to be more homogeneous than heterogeneous such that the communication between similar people occurs more frequently than with dissimilar people \cite{mcpherson2001birds}. The main driving forces for initiating these networks are social influence and homophily. In other words, the importance of establishing connections between people does not rely upon ‘what you know’ but ‘who you know.’ In order to study this phenomenon, various studies have been conducted by sociologists on multiple socio-demographic dimensions of race, age, social class, culture, and ethnicity. For example, friends, colleagues, spouses, and other associates are inclined to mixing with each other who are similar to them than with randomly selected members of the same population  \cite{mcpherson2001birds,yap2015does}. 

Studies in homophily usually have been conducted by surveying a group of human subjects which in most cases belonged in a specific geographical location \cite{cepic2020social,xu2019quantifying,albalawi2019trustworthy,escobar2018better}. For example, one study showed that American high school students have a tendency to make friends with other students that belong to the same race and gender \cite{moody2001race}. Initially, homophily was classified into two categories - status homophily and value homophily \cite{lazarsfeld1954friendship,koiranen2019shared,warren2020building}. Status homophily mainly focused on the social position of the individuals inferring that individuals belonging to similar social conditions are inclined to mixing with one another. Value homophily in contrast is based upon the similarity of thoughts of individuals leading to the belief that individuals with homogeneous thoughts are inclined to connect with others even though differences may lie in their social positions \cite{dehghani2016purity,murase2019structural,dincelli2016information}. Although, researchers have successfully conducted experiments with human beings, the results were often based upon real-world scenarios of only small groups \cite{tang2013exploiting}. In order to fill the gap in the analysis, social media platforms come in handy as social networking sites such as Twitter and Facebook have become extremely widespread, with over 126 million daily Twitter users \cite{pourebrahim2019trip,kassens2019good} and Facebook having approximately 1.2 billion daily users \cite{franz2019using,VanderWeele}. Reactive interfaces like those available through social networks provide users with the opportunity to be more open about their opinions, perspectives, thoughts, likes and dislikes  \cite{liang2018birds,khan2017social}. As a result, social media platforms are becoming more and more popular among users \cite{barreto2017social,mou2017media}. These platforms are known to help users feel more involved. Users feel that they are able to participate in things that are happening around the world. Furthermore, such platforms help users in raising their voice against unjust acts or issues \cite{ghaznavi2015bones}. Therefore, both status and value homophily have been analyzed recently in social networks in order to evaluate whether these types of homophily phenomenon exists in these types of networks. Moreover, if homophily exists, whether it increases or decreases in digital environments has been studied \cite{Nazan,Hywel,basov2019ambivalence}.

The effect of homophily has been vastly studied in different types of social media data. From textual data (Twitter tweets) to follower lists of online social accounts \cite{pan2019twitter}. However, no detailed survey has been conducted to date based on the works of social media networks related to the homophily principle. Therefore, the main aim of this paper is to focus on providing a thorough review of the related works conducted on social media networks based on the homophily principle.

\begin{figure}[h]
    \centering
    \includegraphics[width=17cm]{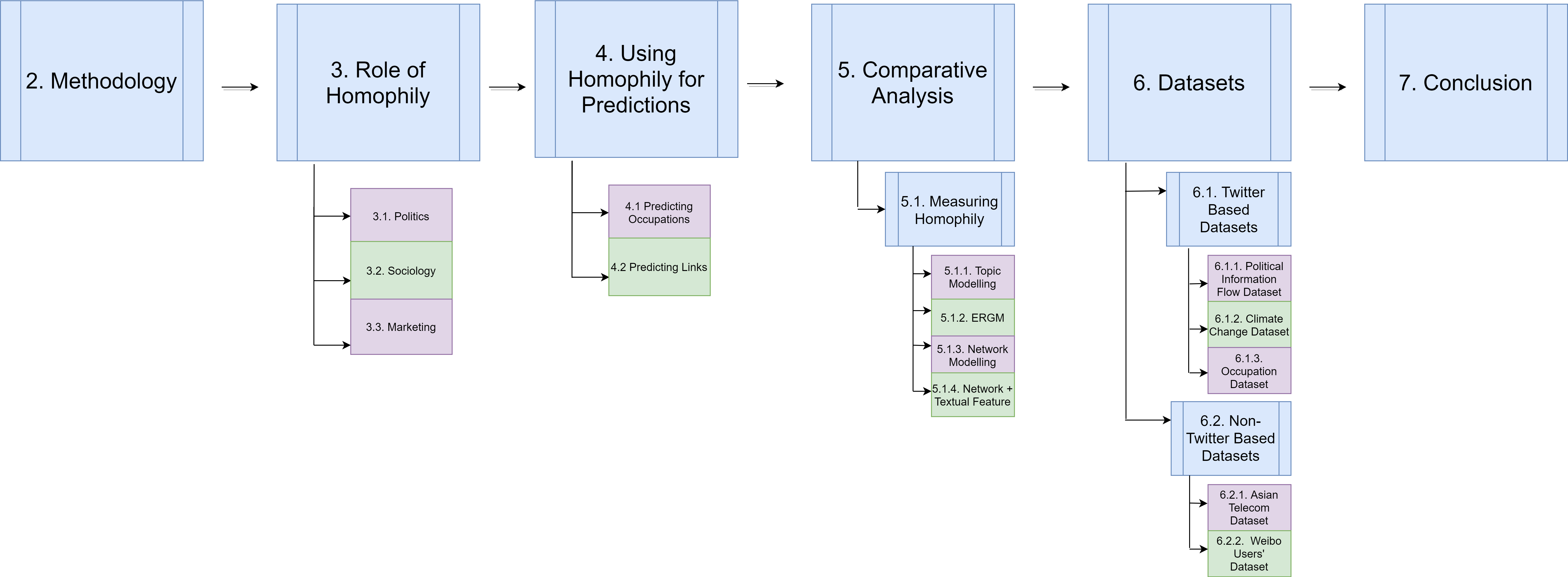}
    \caption{Overall structure of the survey paper}
    \label{Figure:Figure1}
\end{figure}

The rest of the paper is organized as follows. Section \ref{Methodology} presents the methodology that has been used to extract high quality articles in order to conduct the survey. Section \ref{Role} discusses the role of homophily of the various ways in which the homophily effect has been analysed in multiple domains of social media data. Section \ref{Predictions} discusses on the predictions made in many fields of social network by using the homophily effect. Section \ref{Comparative} introduces a comparative study of the social network analysis, conducted by measuring homophily in multiple applications, the different types of network models constructed in each of the proposed models.  Section \ref{Datasets} includes the different types of datasets used to validate these proposed, homophilous models. Section \ref{Conclusion} discusses about the state-of-art methods used for detecting homophily in social networks, the limitations of these approaches and directions for future research. Moreover, Section \ref{Conclusion} draws the conclusion of the survey. Fig. \ref{Figure:Figure1}, shows the overall structure of the paper.

\section{Methodology}\label{Methodology}
\begin{figure}[h]
    \centering
    \includegraphics[width=8cm]{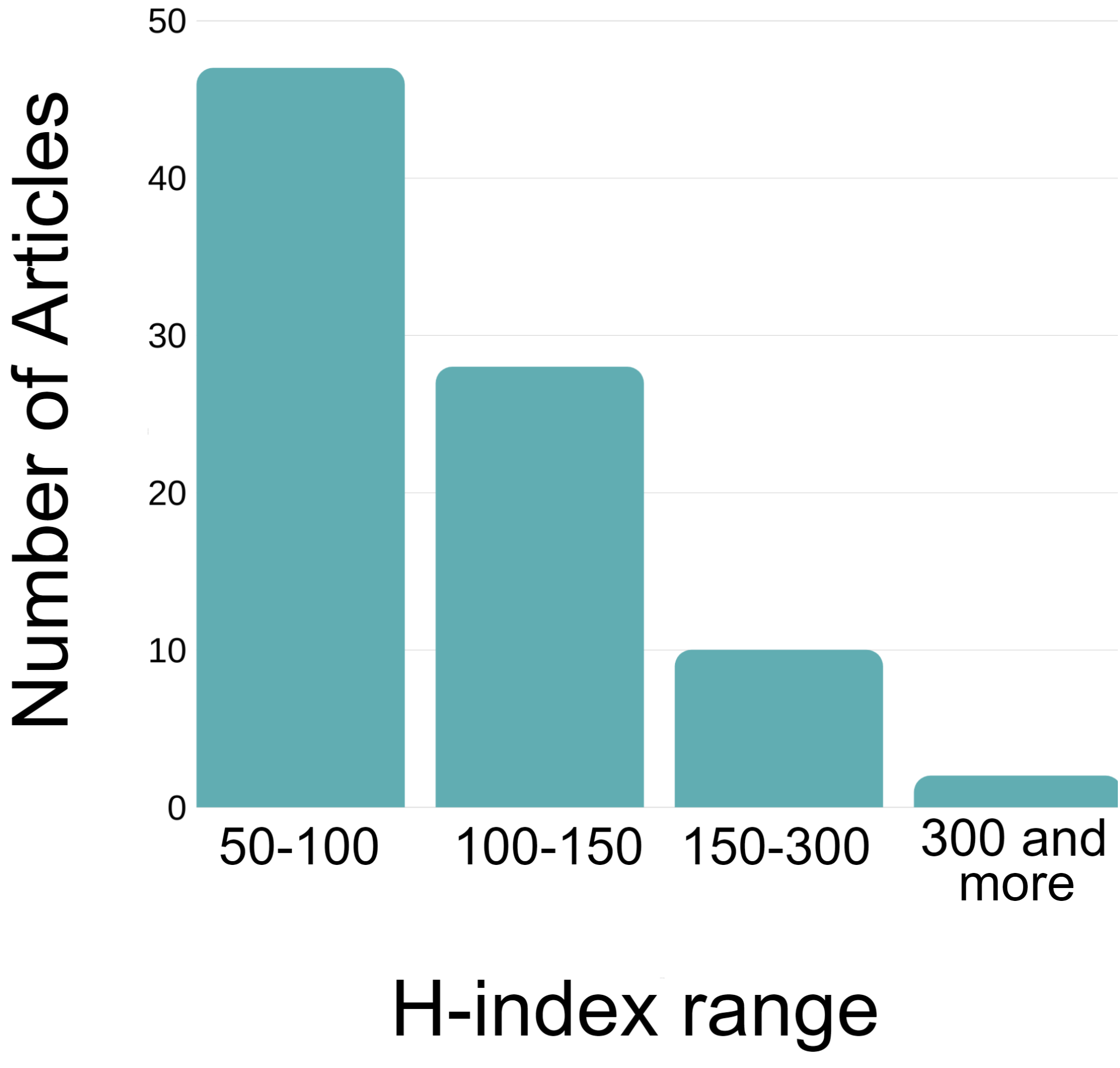}
    \caption{The number of articles' H-index from a range of 50 and above}
    \label{Figure:Figure28}
\end{figure}

Keywords, such as Homophily, social media, and degree distribution have been used to search for papers related to analyzing homophilous models. However, it is not only important to find the appropriate papers based on keywords but also to extract papers from top venues. As such articles have a high impact factor. As a result, the h-index of the venue, where the paper was published and the number of citations of the paper were considered. We have mainly focused on the venues which have an h-index of $50$ or above, from Q1 or Q2 journals, and the articles having a minimum of $100$ citations. Using this information\footnote{\href{https://www.scimagojr.com/journalrank.php}{https:/www.scimagojr.com}}, papers for this survey were obtained by accessing them through university library resources. Recent surveys \cite{litjens2017survey,kamilaris2018deep,goyal2018graph} have also reported adopting the similar approach. Fig. \ref{Figure:Figure28}, shows the h-index of the articles cited in the survey, we can see most of the articles' h-index is from $50$-$100$. As h-index is a venue-level metric which is used to evaluate the impact factor and citations of the publications of the venue. Thus,  Fig. \ref{Figure:Figure28} shows that most of the articles selected have high h-index. The table \ref{table:44} of Appendix \ref{appendix:a} shows the articles selected for this survey with the venue, number of citations, quartile, h-index, as well as the year of publication.

\begin{figure}[htp]
    \centering
    \includegraphics[width=6cm]{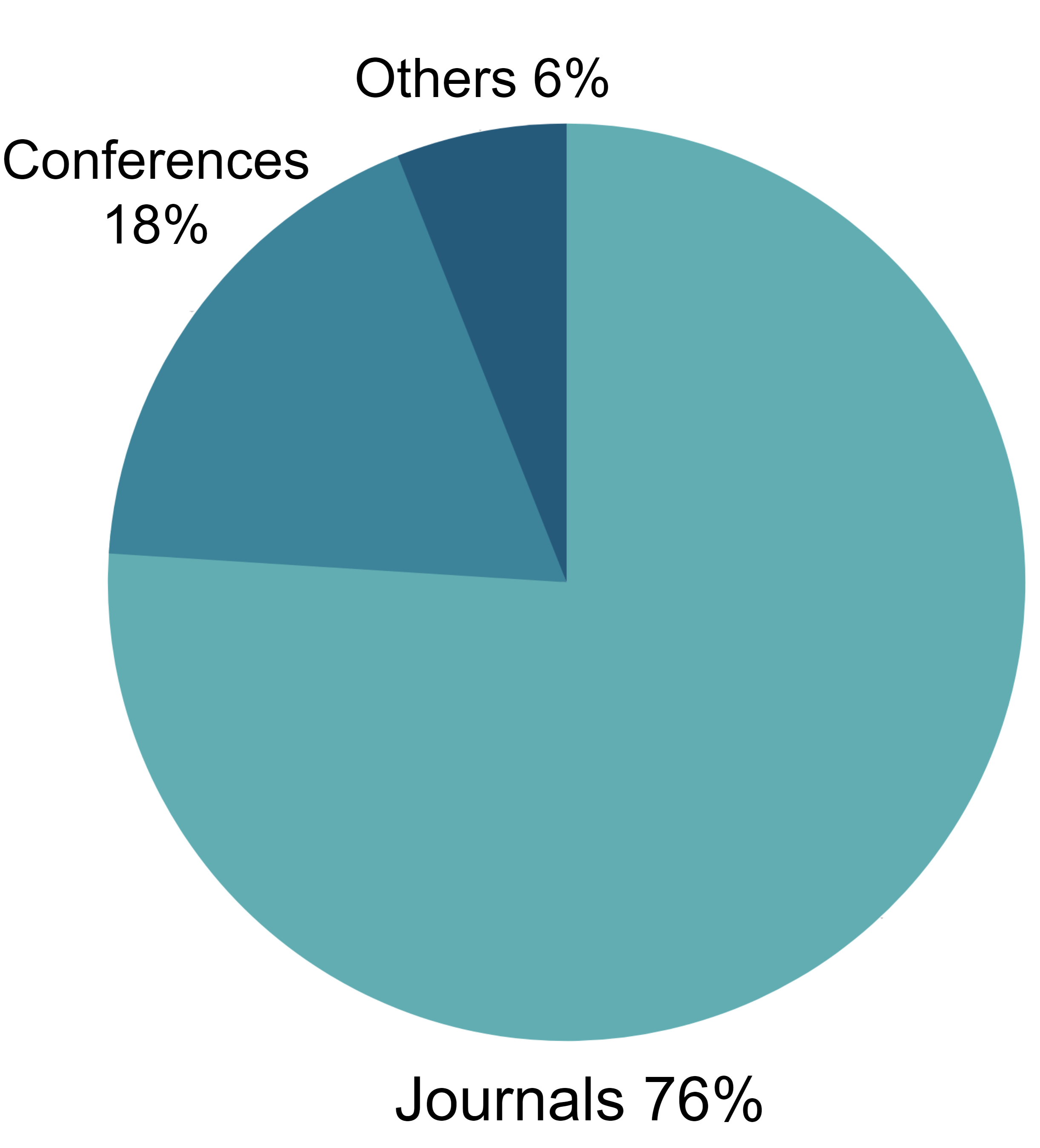}
    \caption{Percentages of Journals, Conferences and other type of articles cited in the survey paper.}
    \label{Figure:Figure88}
\end{figure}

\begin{figure}[h]
    \centering
    \includegraphics[width=6cm]{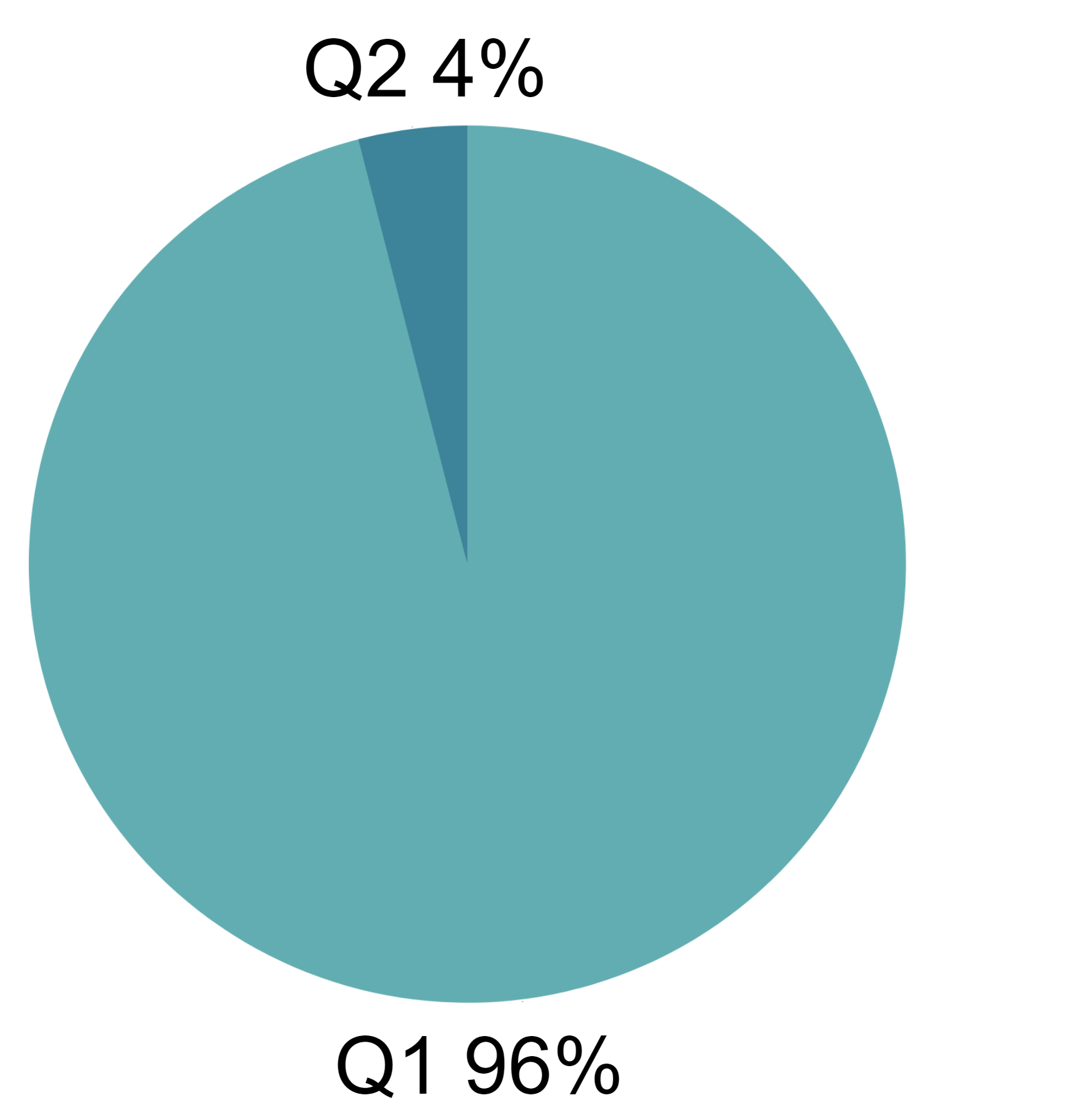}
    \caption{Proportion of Q1 and Q2 Journals referenced in the survey article.}
    \label{Figure:Figure76}
\end{figure}

\begin{figure}[h]
    \centering
    \includegraphics[width=6cm]{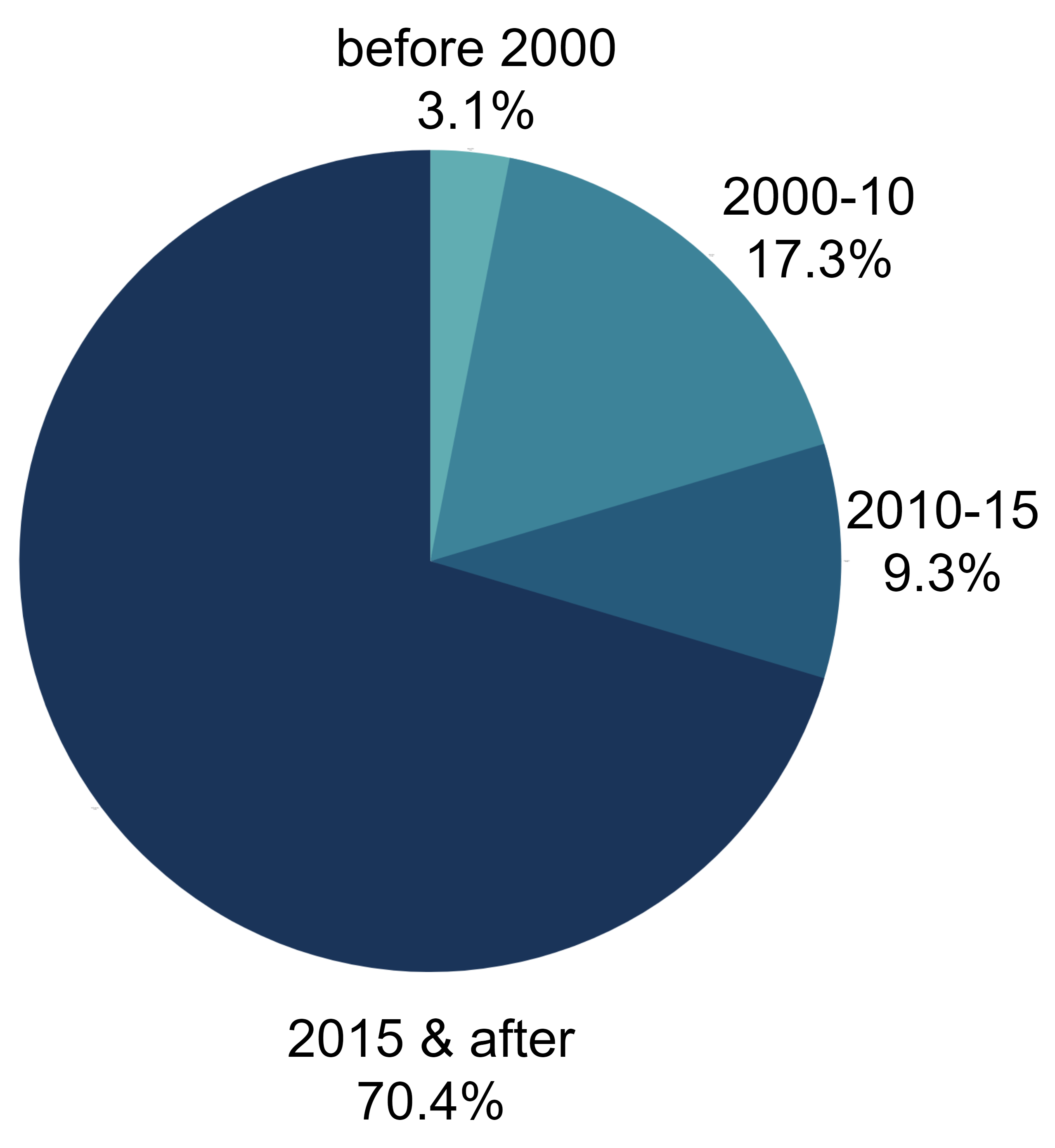}
    \caption{Percentages of the Year of Publication of each of the articles.}
    \label{Figure:Figure25}
\end{figure}

An article from a high h-index venue, with a high number of citations shows that the paper is reliable and trust-worthy for the academic community. Fig. \ref{Figure:Figure88} shows the total percentage of journals, conferences, and other types of articles such as book chapters, workshop papers that have been cited in this paper. It can be seen from Fig. \ref{Figure:Figure88} that most of the articles selected for this survey are from journals. Furthermore, in Fig. \ref{Figure:Figure76} we can see that majority of the articles are taken from Q1 journals. However, for some of the articles, the information about belonging to certain quartile was missing. For such cases, we have only focused on the remaining metrics. Besides, the papers selected were from 2015 onwards so that the approaches used in the recent papers could be studied more exhaustively \cite{li20185g,zhang2019deep}. However, if any articles have major contributions, such as introducing novel algorithms or approaches used in measuring the degree of homophily, then, they are considered for this survey. This is because homophily is not a recent concept and the impact of these papers is more important than the year of publication. Fig. \ref{Figure:Figure25} shows that most of the articles have been selected from 2015 onwards.
\floatstyle{plain}
\restylefloat{figure}

\begin{figure}[h]
    \centering
    \includegraphics[scale=0.3]{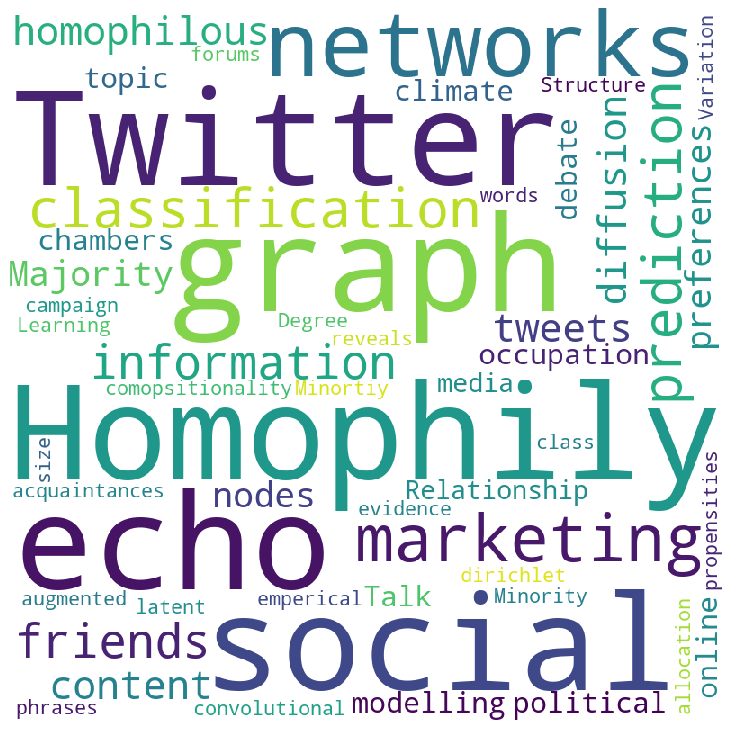}
    \caption{Visual representation of the important words from the abstracts}
    \label{Figure:Figure9}
\end{figure}

A word cloud, as shown in Fig. \ref{Figure:Figure9}, was generated from the abstracts of the papers selected to get a visualization of the most important word in the field of homophily \cite{jin2017development}. We implemented a simple python code to form the word cloud. The abstracts are pre-processed by converting the text to lower case, removing the punctuation, and commonly used English stop words, available in the nltk library. Then the word-cloud is built using a word cloud library. The importance of each word is represented with the font size and color. The darker the color of the word, the more significant the word is. The larger the font size of the word, the more frequently the word has occurred in the abstracts. Fig. \ref{Figure:Figure9}  shows that the homophily word has the largest font which depicts that most of the papers were about homophily related issues. Moreover, the word Twitter and social network has the second-largest font. The color of the font is darker which shows most of the papers discussed about social media. Moreover, the Fig shows that words like social, graph, networks, echo, and homophily are highly co-related to one another. The word \textit{echo} is used, in social media, when users raise their voices and talk about politics or debating about a particular issue. As a result, it can be concluded that most of the abstracts talk about visualizing social network graphs from the issues discussed in social media platforms by using the homophily concept.

\section{Role of Homophily in Social Media}\label{Role}
The Internet has the ability of connecting people, with all kinds of interest, all around the world. As a result, it can be assumed that when social ties are formed between individuals in social media, homophily will be less likely to appear among users. However, homophily has rather shown to increase in social media platforms \cite{zhang2016homophily,boutyline2017social,kwon2017platform,zhang2018equality}. For instance, earliest research has highlighted that, when people used Microsoft Instant Messaging, users were more interested to converse with others belonging to a equivalent ages,location and native language \cite{leskovec2007worldwide}. Moreover, the more individuals communicated with one another, the more related their online searches were \cite{singla2008yes}. Similarly, increased homophily was also detected among groups of Facebook friends, where they had similar thoughts on ideology or political orientation \cite{mayer2008old}. The effect of homophily has been studied in various domains to observe if homophily has any effect on social networks or not. For example, Twitter is a popular micro-blogging platform that has been considered as an effective tool for studying the interactions between the social media users \cite{ladhari2020youtube,mei2019dynamic}. Homophilic studies have been conducted at an exhaustive rate in the domain of politics,  marketing, and sociology as well. Insightful information has been extracted from these approaches. The following subsections discuss in detail the approaches that have been proposed in these domains.

\subsection{Politics}
In the field of politics, homophilic studies range from analyzing the users participating in political debates to observing the network of politically engaged users on a variety of political activities \cite{halberstam2016homophily,Hywel,himelboim2016valence,yaqub2017analysis}. On a politically oriented website named "Essembly", users were observed to form positive and negative ties with people having parallel thoughts and different reasoning on an ideology respectively \cite{hogg2008multiple}. Furthermore, when smaller networks were studied more in-depth, several characteristics such as gender, age and level of education have proven to be strong predictors with network structural characteristics. Moreover, these characteristics were used for investigating the existence and strength of positive ties among individuals \cite{gilbert2009predicting}. 

Homophily principle was also used to study the flow of political information among the majority and minority groups on the Twitter platform \cite{halberstam2016homophily,himelboim2016valence,lai2019stance}. Recent studies have shown that the majority of larger groups received political information more quickly than smaller groups. Both groups were exposed to similar political information and it has been observed that the flow of information was faster among the larger group. Substantial evidence of homophily was detected, when users following a specific political party were more likely to connect with other users following the same party. The flow of information through the social media network was faster among the majority groups since they had more network connections and so they received more information at a faster rate. To sum up, political information is considered as an extremely influential information. Therefore, increasing exposure to such information among the like-minded majority (large) users can further increase political divergence among the users.

Furthermore, social networks of users exchanging views about global warming on Twitter were examined. The users' attitudes towards global warming were classified based on their message content \cite{Hywel,jang2015polarized}. The social networks were categorized by opinion-based homophily and the users were manually labeled as "skeptic" and ``activist" groups based on their message content. Results have shown that, users generally communicate only with other similar-minded users, in communities that are influenced by a common view. Moreover, the messages of like-minded users have shown to be a positive sentiment in most of the cases, whereas, messages from skeptics and activists held a more negative comment. Overall, discussions of climate change in social media often take place in the polarising "echo chambers" where political issues are discussed, and also in ``open forums" and mixed-opinion communities \cite{passe2018homophily,gilbert2009predicting}.

\subsection{Sociology}

Homophily, meaning "love of sameness", is considered to be a sociological theory that like-minded people will be inclined towards each other and will have a tendency to act in a similar way \cite{lazarsfeld1954friendship}. This behavior of individuals has been studied in social media platforms as well. Social media generally consists of majority and minority groups, where the majority group is considered to have stronger connections with one another in its group and also tends to have higher network communications \cite{halberstam2016homophily,basov2019ambivalence}. Compared to the majority group, the minority group not only has fewer members in its group but is also deprived from receiving information quickly \cite{karimi2018homophily}. Thus, the relationship between majority and minority group in social media is studied in depth to observe how the groups and the size of the groups are formed and the groups react to one another in social network \cite{karimi2018homophily,hanks2017status}.

In order to study the influence of homophily between the minority groups, the levels of homophily was calculated by combining the centrality measures with the preferential attachment network \cite{barabasi1999emergence,perra2008spectral,karimi2018homophily}. The model focused on multiple ranges of homophily and density of populations by capturing the degree distributions and ranked the minority groups in empirical social networks of scientific collaboration and dating contacts\cite{karimi2018homophily}. Experimental results have shown that as the volume of the minority group decreased, the heterophilic interactions were greater than the homophilic interactions. However, multiple assumptions were made. For instance, all the members of the minority groups were considered to be equally active and behave in a similar pattern and the group size differences were omitted. These factors can cause a bias estimate in the ranking of the groups. A major drawback was also faced when validating the proposed model such as finding adequate numbers of large scale data representing the minority groups, since, remote and hard-to-reach minority groups are often absent from the social network datasets \cite{shaghaghi2011approaches}. 

\subsection{Marketing}

Comparative analysis has also been conducted on homophily and social influence effects on product purchasing \cite{ma2015latent}. This analysis examined problems related to whether a company should target customers based on homophily or the social influence effect. If a company relies on the homophily principle then they target the existing customer's friends directly as they have a tendency to purchase similar products. Whereas, if the firm emphasizes on social influence of the existing customers then they only target the existing customers and rely on them to promote to their social circles. Therefore, for cost effective marketing strategies, it is extremely important to separate these two effects. However, such approaches are challenging since both phenomenons end up producing similar outcomes. As a result, a product choice model has been designed via the hierarchical Bayesian model which was implemented with a dataset that consists of both communication and product purchase information over a three month period provided by Asian Telecom Company \cite{ma2015latent}. A strong homophily effect was detected on the choice of products. When one of the factors was ignored in the Bayesian model, it resulted in an overestimation of the other factor and this shows that social influence and homophily effects are highly connected with each other. Ignoring any of the factors leads to biased estimates in the Bayesian model. Furthermore, as network structures are versatile in nature \cite{ejima2017metal}, it was difficult for the model to detect strong and weak ties in network structures. This is because, some people in the network might have many friends with weak ties to one another consequently others might have few friends but with extremely strong ties. As a result, the model can be further improved to inspect the strength and the impact of social ties with respect to a customer's decision on product purchasing. This will help to identify customer preferences in such versatile networks. Thus, an improved model is required that can further differentiate the effects of homophily and social influence.

Research was also conducted to study the presence of homophilic patterns based on the usage of hashtags in a Twitter mention network based on a Cause-Related Marketing (CRM) campaign \cite{xu2020hashtag}. CRM is a mutually beneficial collaboration between a corporation and a nonprofit organization. It is designed to promote social responsibility in the public community. However, CRM is a risky and controversial issue since this campaigning varies from receiving skepticism to full support of the customers \cite{barone2000influence,mukherjee2020brand}. Gillette's CRM campaign "The Best Men Can Be" was used, to test the hypothesis that whether homophily exists in such marketing campaigns or not \cite{xu2020hashtag}. The brand's goal was to address concerns based on gender inequality and bullying of men and encourage a better lifestyle for youngsters. The company, moreover, guaranteed to make a donation of $1$ million to NGOs fighting for gender inequality. When the campaign ad was released, the ad received positive feedback from some of the customers because of its positive message \cite{kets2019belief}. However, others felt that the ad was a bit offensive representing men as sexually harassing, and bullying. Thus, it received negative feedback from the rest of the customers \cite{van2019adaptive}. Hence, two groups were formed in the social media where one group was supporting the cause while the other group opposed its motive. As the brand is well-known and discussions on this campaign became a trending topic on Twitter. Thus, this CRM campaign was an ideal fit to analyze how the users communicated and reacted with other users on the Gillette's ad. For CRM's marketing campaign, topic modeling was used for extracting information \cite{papadimitriou2000latent}. Topic modeling was conducted on 100,000 original tweets, profiling the topics related to the CRM campaign tweets \cite{blei2003latent}. Based on the users’ engagement, the network of the CRM event's related hashtags were analyzed with the aid of Exponential Random Graph Models (ERGMs) \cite{snijders2002markov}. ERGMs are statistical models that are commonly used for analyzing data regarding online social networks \cite{stivala2020exponential,colladon2019measuring}. Results generated from this model showed an increased tendency of homophily on the network of users. The degree of homophily inspected was based upon the common views of the users. The results of topic modeling on Twitter have revealed that users are highly dependent on established social networking platforms to discuss important issues \cite{kets2019belief,grace2018friend}. Furthermore, users tend to react more to the tweets of influential, popular users which enable the users to be more reactive during online discussions. Moreover, these users showed homophily on the usage of hashtags. Thus, ideological hashtags served as measures of homophily as ideological hashtags refers to a person's identity and thoughts \cite{blevins2019tweeting}. One such example of ideological hashtag is the usage of \textit{\#BlackLivesMatter} or \textit{\#AllLivesMatter} which reflects the user's ideological position is based on the social justice issues to a large extent. Therefore, hashtags not only express a user's self-identity but also helps similar users to identify and connect in a versatile community \cite{van2019adaptive,an2016greysanatomy}.

\section{Using Homophily for Predictions }\label{Predictions}
Homophily concepts can be implemented for predicting certain features. For example, homophily principles have been applied in the link prediction area for studying the probability of one user to be connected with another user. On the other hand, homophily in Twitter has also been examined to predict occupation of users based on the information of the users' followers and followings' IDs. The usage of homophily concept in predicting certain features are discussed below.

\subsection{Predicting occupations}

In order to predict the occupation of Twitter users, the homophily principle was used to conduct social network analysis on the biographical content of the user's follower/following community \cite{pan2019twitter}. Occupation prediction is considered as a multi-classification problem since the model is specialized in predicting multiple occupational classes. Furthermore, the results concluded that a user's follower/following community provides insightful information for identifying the occupational group of each of the users. The model was designed with Graph Convolutional Network (GCN) \cite{kipf2016semi} which has enhanced the model to work efficiently by training on only a small fraction of data. Graph convolutional network is a recently proposed graph based neural network learning model, which specializes on learning graph-structured network data \cite{kipf2016semi}. Thus, by using the homophily principle and GCN, a better result was achieved for predicting occupation class with an accuracy of 61\%. A similar work was done to predict the occupational class of Twitter users where the  dataset contained the historical tweets of the users \cite{preoctiuc2015analysis}, however, an accuracy of only 50\% was achieved.

\subsection{Predicting Links}
The homophily principle was used as motivation in various works of link prediction, where link prediction is calculated based on the similarity between two entities.As a result, it can be used to predict future possible links in social networking platforms \cite{currarini2016simple}. For example, researchers have proposed a model to investigate the associated links of document's topic distribution between people discussing about related topics. This study shows how topic distribution is mainly affected by the distribution of the topics of its nearest neighbors \cite{yang2015birds,kim2017effect,xu2020hashtag}.

Particularly, a joint model was proposed in which link structure has been applied to define clusters. Here, each of the clusters was allocated with its own segregated Dirichlet prior for topic distribution. Using such priors have shown to be very helpful as in previous works only document priors were applied \cite{zhu2012medlda,mimno2012topic}. Discriminative and max-margin approaches \cite{zhu2012medlda,zhu2014gibbs} have been used for designing the contextual documents and generating good link predictions. Moreover, lexical terms have been used in the decision function in order to improve the strength of the prediction \cite{nguyen2013lexical}.

In summary, users in social media are not only comfortable at expressing their self identities but also have a tendency to connect with one another, with similar interests, in a versatile community. Several studies have been carried out to study the homophilic patterns especially among the majority and minority groups. Studies have deduced that the majority and larger groups receive information faster than smaller, minority groups and information reaches like-minded individuals more quickly \cite{mahmood2017will,huber2017political,halberstam2016homophily}. Whereas, the minority groups are deprived from receiving information instantaneously as such groups have fewer members in their groups hence have fewer network connections \cite{karimi2018homophily}. On the other hand, a strong homophily effect was detected on customers having similar product tastes and based on the attraction of users having a common view \cite{ma2015latent,Hywel,xu2020hashtag}.

\section{Comparative Study of Related Works for homophily detection}\label{Comparative}
In this section, comparative analysis has been performed on various approaches proposed in order to detect and calculate the level of homophily. In addition, the network and language models defined by each of the state of art methods are also explained in details. 

\subsection{Measuring the degree of Homophily}
Multiple approaches have been proposed for measuring the level of homophily with the aid of topic modeling and network modeling. Recently, the degree of homophily has also been calculated by combining both textual as well as network features by using a highly efficient neural network model that has outperformed the existing traditional methods \cite{pan2019twitter,karimi2018homophily,halberstam2016homophily,xu2020hashtag}.

\subsubsection{Topic modeling}
Topic modeling is considered as an unsupervised machine learning technique which does not need the aid of humans to determine the topic of a set of documents \cite{belford2018stability}. It can automatically detect the main theme of given paragraphs or documents by clustering groups of words or similar expressions that best portrays the set of documents or paragraphs. The topic model proposed based on the homophily concept specializes in detecting high-quality topics to test the hypothesis that whether people talking about similar topics are connected with each other or not \cite{yang2015birds}. For example, the Latent Dirichlet Model (LDA), is a topic model, that maps the documents to the topics based on the distribution of words \cite{blei2003latent}. LDA model was modified with the concepts of homophily to not only detect the high-quality topics but also predict whether the people having similar posts in social media are connected with one another or not \cite{blei2003latent}.

Generally, the most frequent words of each of the documents are aggregated into \textit{K} words clusters by using the \textit{k} means algorithm \cite{macqueen1967some}. Thus, for any word token $w_{d,n}$, for the word token belonging to a cluster \textit{k}, any other token $z_{d,n}$ being a neighbor of $w_{d,n}$ will also belong to cluster \textit{k}. The $d$ represents the document and $n$ represents the number of documents. Therefore, in order to find the topic $k$'s major words, skip-gram transition probability \cite{mikolov2013distributed} is calculated for each $w_{k,i}$ word as in Equation \ref{first_equ}.

\begin{equation}\label{first_equ}
S_{k,i} = \sum\limits_{j=1,j \neq 1}^{N_k} p(w_{k,j}|w_{k,i}) 
\end{equation}

where, $N_k$ indicates the number of words in topic \textit{k},  words with the highest probabilities are used as the designated topic words for each of the documents in the sample. Regression is used to compute the topic distribution between $d$ and $d^\prime$, for predicting the link between the two documents, which is depended on the similarity of their topic patterns.
Therefore, the regression value is defined in Equation \ref{second_equ}.

\begin{equation}\label{second_equ}
R_{d,d^\prime} = \eta^T(\overline{z_d} \circ \overline{z_{d^\prime}}) +  \tau^T(\overline{w_d} \circ \overline{w_{d^\prime}})
\end{equation}

where, $\overline{z_d} = \frac{1}{N_d}\sum\nolimits_{n} z_{d,n}$. Similarly, $\overline{w_d} = \frac{1}{N_d}\sum\nolimits_{n} w_{d,n}$; $\circ$ denotes the Hadamard product \cite{horn1990hadamard}; $\eta$ and $\tau$ are the assigned topic based weight vectors and document link predictions.

In some studies, the perplexity metric was used as a measurement for evaluating the model's topic modeling performance \cite{gamallo2017perplexity,boldsen2019identifying,conneau2019cross}. Perplexity is a measurement based on the quality of a probability model predicting a sample. Results have assured that the proposed model outperformed the traditional LDA model for topic modeling. Furthermore, for validating the model in terms of its performance for predicting document link prediction, the Predictive Link Rank (PLR) metric was used. PLR outputs the average rank of a document with the documents to which it has been linked with. High training performance was achieved that showed user interactions can contribute to better link prediction. However, the testing performance score was much lower than the training performance score. This shows that the model has over-fitted since the model could not perform well with the testing dataset. Even though the new model outperformed the traditional LDA method, for document link predicting task the overfitting issue was not resolved.

LDA-based topic modeling focuses on topics co-occurring frequently. However, the main drawbacks of LDA based approach is the need of specifying the "appropriate" number of topics that the LDA has to predict \cite{blei2003latent}. Statistical indices have been proposed to address this issue \cite{arun2010finding,cao2009density,deveaud2014accurate} which includes differentiating each pair of the topics by the difference of each pair of topics or their distance. Although, these methods can approximately calculate the number of topics in a given corpus, proper gold standards or benchmark still does not exist. Therefore, human interpretation is still required  for rendering the topics into the unsupervised topic modeling method \cite{chang2009reading}. Most importantly, the performance of these methods are not analyzed in documents such as a tweet which consists of only sentences with few number of words. Thus, for measuring the homophily of users using similar hashtags in Twitter posts, the co-occurrence of the topic was calculated by using Mimno \textit{et al.}'s approach \cite{xu2020hashtag,mimno2011optimizing}. Mimno \textit{et al.} state that for any document, the leading words in a topic profile are likely in the same document. The coherence in Mimno \textit{et al.}'s approach is defined as:

\begin{equation}\label{eigth_equ}
C(t;v^{(t)}) = \sum\limits_{m=2}^{M}\sum\limits_{l=1}^{m-1}\log\frac{D(v_m^{(t)},v_l^{(t)})+1 }{D(v_l^{(t)})}
\end{equation}

In Equation \ref{eigth_equ},  $D(v)$ denotes the frequency the document, word $v$ and $V^{(t)}$, topic $t$ has a list of $M$ words.The topic model can predict better when the result has an output that is close to zero. \cite{puranam2017effect}. In order to evaluate the quality of the predicted topics, the distance between each of the topic was calculated using Jensen-Shannon divergence \cite{puranam2017effect,steyvers2007probabilistic}.

\subsubsection{Exponential random graph modeling} 

Exponential random graph models were used to evaluate whether the networks of users engaged in conversations influence the users' response in online discussions \cite{xu2020hashtag}. In ERGMs, nodes' connections at the same degree is considered to be an indicator of users conversing frequently \cite{morris2008specification}. These graph models are statistical models that are composed of network structures \cite{lusher2013exponential}. ERGMs have been used in multiple domains of the social networks such as social media settings and to study communication between the users. For modeling the presence of any ties that may exist between the network's local and structural factors \cite{gonzalez2009opening,robins2007introduction,saffer2018reconsidering}.

Generally, the ERGM model is designed by aggregated tweets and hashtags which are posted by the users. \cite{xu2020hashtag}. The top hashtags are categorized as either conceptual or ideological markers based on the definitions  provided by Blevins \textit{et al.} \cite{blevins2019tweeting}. Ideological hashtag refers to the identity or identification, perspectives. On the other hand, conceptual markers are considered as personal thoughts on particular events.  Then, the users mentioned in the network as @user and the classified hashtags are fed into the statistical modeling as nodal attributes. The mention network is modeled because it helps to visualize a discussion that starts by actively exchanging information with one another rather than relying on what others have said.

\subsubsection{Network modeling}
Network modeling is a  flexible way of representing a group of connected objects. The objects are represented as nodes or vertices and the connection between the nodes are represented by edges. Network modeling is generally used to visualize the various types of networks. These models are constructed from the social media data consisting of users and how the users are connected. Many researchers have claimed that the flow of information in social media is dependent highly on the majority groups where more users are connected. Moreover, the majority groups have a larger social circle compared to the minority groups. The reason behind having a larger social circle is due to homophily \cite{halberstam2016homophily,karimi2018homophily}.

The homophily among majority groups was studied between a network of politically engaged users of Twitter \cite{halberstam2016homophily}. Due to the shortage of measuring the political orientation and ideology of the Twitter users directly. The research emphasized on the users following politicians from the two major parties- \textit{Conservatives} and \textit{Liberals} of the House of Representatives, in the 2012 general election. In order to analyze the degree of homophily, the individuals are divided into two groups - conservatives \textit{(C)} and liberals \textit{(L)}, depending on which political party each group supports such that: $t \in { C, L }$. The group sizes are normalized and symbolized as  \textit{wt} which is the weight of the tweet, where, $ w_C + w_L = 1 $. Conservatives was randomly selected as the majority group and liberals was considered as the minority group. Therefore, $w_C \geq 0.5 $.

When two individuals belonging to the same group are randomly selected, then the probability of the two individuals communicating with each other is denoted by $\pi_s$ and the probability of the two individuals connecting with each other belonging to different groups is represented by $\pi_d$. Moreover, it is also logical to assume that individuals belonging to the same group have a higher tendency to interact with each other compared to two individuals of two different groups communicating with one another. As a result, $\pi_s > \pi_d$. Thus, an individual belonging to a group \textit{t} will be having similar $\pi_sw_t$ interactions, and  $\pi_d(1-w_t)$ different interactions.
Therefore, in this study, the Homophily principle has been evaluated as such:
\begin{equation}\label{fifth_equ}
H_t = \frac{\pi_sw_t}{\pi_sw_t+\pi_d(1-w_t)}
\end{equation}
In Equation \ref{fifth_equ}, the greater the value of  $\pi_sw_t$, the higher the degree of homophily. As a result, if conservatives are more prominent and links are formed with similar types, then conservatives would tend to be more homophilous in nature. Similarly, the liberals would be more heterophilous. Furthermore, the time taken to reach the information to majority and minority groups were also taken into account. So it was considered that each user produces information with probability of $\varepsilon$ at time $ \tau = 0 $. Bass model or Bass Diffusion Model was used in order to generate the proposed model \cite{bass1994bass}. The Bass model uses differential equation that describes the procedure of new products getting adopted in a population. In this case, the product is the political tweets. If interaction occurs between two users, then the user exposed to information transfers the information to unexposed user with a probability of \textit{q}. Hence, by following the Bass model, the rate of information diffusion is defined as such: 
\begin{equation}\label{sixth_equ}
F^{\tau}_t = F^{\tau - 1}_t + (1-F^{\tau - 1}_t) f^{\tau}_t
\end{equation}

In Equation \ref{sixth_equ}, $F^{\tau}_t$ is defined as the fraction of group \textit{t} receiving information at time $\tau$ which is then connected to the fraction transmitted information at time $\tau - 1$. $F^{\tau}_t$ is the chances of group \textit{t} getting information at time $\tau$ if not being exposed to information at time $\tau - 1$. Therefore, $f^{\tau}_t$ is defined as:
\begin{align}\label{seventh_equ}
f^{\tau}_t = qw_t\pi_s F^{\tau - 1}_t + q(1-w_t)\pi_d F^{\tau - 1}_{-t} - q^2w_t(1-w_t)\pi_s\pi_d F^{\tau - 1}_{t}F^{\tau - 1}_{-t}
\end{align}

The symbol $-t$ in Equation \ref{seventh_equ} refers to the other group. The first term denotes the likelihood of receiving the information from the individual belonging to the same group. Second term denotes the likelihood of receiving the information from a different group. Third term refers to the likelihood of both groups receiving information. In conclusion, if biased interactions are present, ($\pi_s > \pi_d$), the majority group member will receive information faster than the minority group ($F^{\tau}_C > F^{\tau}_L$) for every $\tau$ times. Based on receiving a higher probability score for the majority group, it is deduced that homophily is directly proportional to the rate of flow of information among the users. Moreover, the diffusion of information is relatively uniform with groups having a higher number of connections based on having similar political orientations. Thus, larger groups are exposed to information at a faster rate. Therefore, a close relation of homophily and diffusion of information is shown in this approach.

A similar approach was also used to measure the level of homophily between the minority groups \cite{karimi2018homophily} which is shown in Equation \ref{fifth_equ}. The homophily was used as an additional parameter in the famous model of preferential attachment proposed by Barabási and Albert \cite{barabasi1999emergence}. Preferential attachment means that a node is more likely to receive new links if it has higher number of connections.
Thus, such nodes are more powerful since they can tightly hold links with one another. The growth of complex evolving networks was calculated using the fitness model which is based on the Barabási–Albert model \cite{karimi2018homophily}. In this model, nodes with different type of characteristics can grasp links at different rates. Hence, the fitness is calculated by the degree distribution of each of the nodes. This is how the model can predict a node's growth. In the Preferential Attachment model, at each step, a new node which just approached, its degree and group attachment is calculated for the possibility of the node to be attached to the pre-existing nodes. The chances of node $j$ to be connected to node $i$ is defined as:

\begin{equation}\label{twenty_equ}
\Pi_i = \frac{h_{ij}k_i}{\sum_l h_{lj}k_l}
\end{equation}

In Equation \ref{twenty_equ}, $k_i$ generally, represents the degree of node $i$ and $h_{ij}$ is the similarity between nodes $i$ and $j$. The similarity between each of the nodes is build, based on the nodes' attachment when the network was generated. If by any chance, the new node is not confronting individuals from the same network, it can stay deserted till the node confronts a newly approached node that is coming from the same network.

On the other hand, when the online debate on climate change was studied, the degree of homophily among the individuals was measured on the number of times the edges were connecting users on homogeneous/heterogeneous views\cite{Hywel}. The high frequency of edges between the homogeneous users, and similarly, the low frequency of edges between the heterogeneous users were considered as the measure of homophily. The probability of picking node $i$ as the root or focus node for a given edge, were denoted by:

\begin{equation}\label{twentyone_equ}
P_{source} (i) = \frac{k_{out}(i)}{\sum_{j\in a,s} k_{out}(j)}
\end{equation}

\begin{equation}\label{twentytwo_equ}
P_{target} (i) = \frac{k_{in}(i)}{\sum_{j\in a,s} k_{in}(j)}
\end{equation}

In Equations \ref{twentyone_equ} and \ref{twentytwo_equ}, $k_{out}$ is node out-degree and $k_{in}$ is node in-degree. This mathematical technique generates nodes' networks with homogeneous degree distributions.

\subsubsection{Combining Network and Textual features}

Recently, a new neural network model has been proposed known as the Graph Convolutional network (GCN) \cite{kipf2016semi}. GCN has been used for extracting the textual and the network features for identifying the homophily connection between the Twitter users and their followers/followings list \cite{pan2019twitter}. GCN has not only enabled the model to achieve a high performance, but also the model was successfully trained  with only a fraction of data. GCN graph-based neural network model \textit{f(X,A)} with layer-wise propagation rules is defined as such: 
\begin{equation}\label{fourth_equ}
\widehat{A} = D^{-1/2}(A + \lambda I)D^{-1/2}
\end{equation}

\begin{equation}\label{fifths_equ}
X^{l+1} = \sigma(\widehat{A}X^{l}W^{l}+b^{l})
\end{equation}

In Equations \ref{fourth_equ} and \ref{fifths_equ}, $X$ denotes the matrix of the features for each of the nodes(users). $X^{0}$ is the initial feature with a input size of \textit{(nodes * features)} and $A$ is the adjacency matrix of the dimensions \textit{(nodes * nodes)}. $D$ represents the degree matrix of $ A + \lambda I $, where $\lambda$ is the hyperparameter that controls the weight of the node among its neighbor hood. $W^l$ , $b^l$ are the trainable weights and bias for the $l^th$ layer. In each GCN layer, nodes accumulates its closest neighbors' features by linearly converting the representation using weight \textit{(W)} and bias $b$ respectively. $\sigma$ represents the activation function used in the GCN model for optimizing the performance. Then, the number of GCN layers determines the path of the node from its closest neighbors' features.

The inputs of the adjacency matrix $\widehat{A}$ are all the network IDs (target users and their followers and following list IDs). A feature matrix of the biographical descriptions of the each of the target users' followers/following lists. This is because the biographical description of each of the target users' followers/following lists might be similar with the target users. Pan et el. claimed that an accuracy of 61\% was obtained, which outperformed the results of the existing methods \cite{pan2019twitter}. Thus, GCN was able to extract the rich network and textual information in order to learn the homophily connection between the users. However, the model was trained with only a small fraction of data and thus if a larger amount of data would be used the model would achieve a better result for predicting the occupation of target users. 

In summary, various types of models have been proposed for measuring the degree of homophily. Mainly network modeling and topic modeling have been used to find out the strength of a relationship of a user with the user's nearest neighbors. To also investigate, how many users are involved with one another. Network modeling focused on the network features which involved calculating the number of connections each user has and the strength of the connection of the users with one another. On the other hand, topic modeling focused on clustering the main topics of each of the users, based on their textual tweets and how similar the  usage of topics are with one another. Recently, the degree of homophily has also been evaluated by incorporating textual and network features with highly efficient neural network model that has outperformed the existing traditional methods for multi-classification problems.

\section{Datasets}\label{Datasets}
Multiple datasets have been used for measuring the degree of homophily in multiple fields such as link prediction and flow of information among majority and minority groups  \cite{pan2019twitter,halberstam2016homophily}. For example, the dataset provided by Asian Telecom Company was used to analyze if the homophily effect can influence product purchasing \cite{ma2015latent}. This section discusses the various datasets used to study the homophily effect.

    \newcolumntype{C}[1]{>{\centering\arraybackslash}m{#1}}
    \begin{sidewaystable}
    \vspace*{17cm}
    \scriptsize
    \setlength\tabcolsep{3pt}
        \normalsize
        \begin{tabular}{|p{0.5cm}|p{1.5cm}|p{1.2cm}|p{3cm}|p{1cm}|p{0.8cm}|p{4.1cm}|p{6cm}|}
        \rowcolor[HTML]{343434}
         {\color[HTML]{FFFFFF}} &
         {\color[HTML]{FFFFFF} \textbf{Dataset}} & {\color[HTML]{FFFFFF} \textbf{Source}} & {\color[HTML]{FFFFFF} \textbf{Size}} & {\color[HTML]{FFFFFF} \textbf{Public \newline Private }}& {\color[HTML]{FFFFFF} \textbf{Year}} &
         {\color[HTML]{FFFFFF} \textbf{Advantage}} &
         {\color[HTML]{FFFFFF} \textbf{Additional \newline Information}} \\
    1 &Occupation Dataset &Twitter & 34,630 unique users \newline and 586,303 edges & Public & 2019 & Achieved the highest accuracy \newline at predicting the \newline occupation of 5000 target users \newline & Predicted the occupation of target users based on the biographical content of the target users’ social circles \cite{pan2019twitter}.\\
        \hline
        2 &Political Information Flow Dataset  & Twitter & 2.2 million Twitter users \newline with 90 million network links & Private & 2018 & Time taken for information \newline to flow among the majority of voters could be efficiently calculated & Information flows faster among users \newline with higher number of connection compare \newline to users with less number of connections \cite{halberstam2016homophily}. \\
        \hline
        3 &Majority-Minority Dataset & Generate artificial undirected network & 
        The undirected network consisted of \newline 5000 nodes and averaged over 20 simulations & Private & 2018 & Calculates homophily by combining the centrality measures with the preferential attachment network & Captures the degree distributions \newline and ranks of the majority and minority in empirical social networks \cite{karimi2018homophily}.\\
        \hline
        4 & Climate Change Dataset  & Twitter &  590,608 distinct tweets from 179,180 distinct users was used & Private & 2015 & Hashtag analysis was conducted using mean Sorensen similarity \cite{sorensen1948method} & Sorensen similarity could successfully detect homophily as greater values showed a greater constancy among a major population of active users \cite{Hywel}. \\
        \hline
        5 &Emotions and Political Talk Dataset  & Twitter & 70 datasets were collected based on 10 controversial topics, each dataset has tweets of 1500 users & Private & 2016 & Identifying the emotions based on political conversation by using k-mean clustering & Their findings suggested that oppositional tone were associated more with negative emotion clusters, while supportive clusters overlapped more often with positive emotion clusters \cite{himelboim2016valence}.\\
        \hline
        6 &Debate Dataset  & Twitter & Collection of 900,000 tweets & Private & 2019 & Multi-classification problem of detecting arguments between users by training the model with labels - against, favor or none & Achieved f1 score of 0.60 by using Linear SVM method for predicting any argument occurring between users \cite{lai2019stance}.\\
        \hline
        7 &Asian Telecom Dataset  & Asian Telephone company & 300 million phone call histories of the company's approx 3.7 million customers & Private & 2015 & Hierarchical Bayesian model was developed with the communication information & Strong homophily effect was detected on product purchasing \cite{ma2015latent}.\\
        \hline
        8 &Weibo Users’ Dataset & Sina Weibo website & Posts of 2000 users & Private & 2015 & Homophily is calculated  by predicting links between the users’ posts & The model could obtain better link prediction scores between the users by calculating the rate of similarity of each of the tweets \cite{yang2015birds}. \\
 \hline
 9 & CRM Campaigning Dataset & Twitter & 100,000 posts from 75,302 unique twitter users & Private  & 2020 & performed topic modeling on original tweets by using exponential random graph models (ERGM) & Strong homophily was detected among users using certain hashtags \cite{xu2020hashtag}.\\
\hline
    \end{tabular}
\caption{Details of some of the datasets used to validate the presence of Homophily in the digital environment}
    \label{table:33}
    \end{sidewaystable}

\subsection{Twitter-Based Datasets}
Currently, there are no standard datasets that have been used to study the effect of homophily. Rather in most of the cases, datasets were generated with the aid of Twitter Search API \cite{o2010tweetmotif}. Relevant words or hashtags have been used to find tweets and hashtags among the users who have posted such tweets. For example, to search for tweets about climate change, hashtags such as  \textit{\#climatechange} and \textit{\#globalwarming} are used to find such specific tweets about climate change along with the user information \cite{Hywel}. Based on this search, an extensive network is generated from these users. Different approaches such as network modeling and topic modeling are used to measure the degree of homophily. In Table \ref{table:33}:$ 1, 2, 4, 5, 6, 9$ show examples of some of the datasets that were generated to study the effect of homophily by using this approach, as the mode of data collection was very much alike. Hence, the description of the following datasets gives an overview of how networks are generated by using the Twitter Search API.

\subsubsection{ Political Information Flow Dataset}
The twitter dataset for analyzing the flow of political information among majority and minority groups was constructed by targeting the politically engaged users \cite{halberstam2016homophily}. These users were following at-least one account of a candidate running for the 2012 US elections. As a result, over 2.2 million users’ data were collected from which 90 million network links were approximately identified. Users following more accounts of Republican political candidates than accounts connected with Democrats candidates were categorized as conservatives. Similarly, the users following more Democratic accounts were categorized as liberals. To measure the level of communication among the groups of supporters, approximately 500,000 retweets of the candidates’ tweets and tweets that mention candidates were also collected and analyzed. The flow of political information among the groups of Twitter users was measured based on whether or not the users received a candidate tweet or mention through these networks. Moreover, the rate of information flowing through the political network was taken into account by measuring the time taken for these retweets to diffuse across the networks.

\subsubsection{Climate Change Dataset} 
Twitter API was used to collect tweets between January 2013 and May 2013 that consisted of the trending hashtags on global warming such as- \textit{\#globalwarming}, \textit{\#climatechange}, \textit{\#agw (an acronym for “anthropogenic global warming”)}, \textit{\#climate}, and \textit{\#climaterealists} \cite{Hywel}. Moreover, followers of each of the users posting such tweets were also identified. Hence, $590,608$ distinct tweets from 179,180 distinct users were used to generate the dataset. Hashtags were mainly used to search tweets as Twitter users commonly use hashtags to pinpoint a specific occasion. This enables users to search and participate in a relevant discussion. Mean Sorensen similarity also known as F1 score, \cite{sorensen1948method} was calculated for each of the hashtags. As Sorensen similarity score is within a range of $0$ (no overlap) to $1$ (identical). Greater values showed greater constancy among a major population of active users.

\subsubsection{ Occupational Twitter Dataset } 
While most of the datasets were generated by using relevant keywords/hashtags, the occupation dataset was generated by using the biographical content of each of the target users. The Occupation dataset of Table \ref{table:33} shows the details of the dataset. Occupational Twitter Dataset has public access and maps $5,191$ Twitter users to $9$ major occupational classes \cite{preoctiuc2015analysis}. The dataset consists of User IDs and the historical tweets of each of the users. The Occupational prediction problem is considered as a multi-classification problem as the model focuses on predicting multiple occupational classes. The occupations of each of the users were manually labeled with the aid of Standard Occupation Classification (SOC) from the UK \footnote{\href{www.ons.gov.uk}{https://www.ons.gov.uk/}} which is shown in detail in Table \ref{table:4}.
 The dataset was initially used to predict the occupation of the main Twitter user based on their historical tweets \cite{preoctiuc2015analysis}. Later, the dataset was further extended, to analyze deeper into the network information. The biographical descriptions of the following and followers' IDs for each of the main user ID \cite{pan2019twitter} were added. Biographical descriptions were extracted from the $160$-character-long summary that a user writes about themselves in their profile. Thus, the extended dataset had the followers and followings information for about $4,557$ main users. Due to account suspension and protected tweets, the remaining users account could not be reached. Table \ref{table:4} shows the occupational class distribution of the users' occupational dataset. The biographical information of the main users were not taken into account since the main users' occupations were manually annotated in the dataset.

\begin{table}[ht]
    
    \begin{tabularx}{\columnwidth}{|s|b|s|}
        \hline
        \rowcolor[HTML]{343434}
         {\color[HTML]{FFFFFF} \textbf{Occupational class}} & {\color[HTML]{FFFFFF} \textbf{Standarad Occupation Classification}}  & {\color[HTML]{FFFFFF} \textbf{Users}}   \\
        \hline
        1 & Managers, Directors, Senior Officials & 461\\
        \hline
        2  & Professional Occ. & 1,611 \\
        \hline
        3 &  Associate Profess., Technical Occ. & 926   \\
        \hline
        4 & Administrative Secretarial Occ. & 162   \\
        \hline
        5 & Skilled Trades Occ. & 768   \\
        \hline
        6 & Caring, Leisure, Other Service Occ. & 259   \\
        \hline
        7 & Sales and Customer Service Occ. & 58   \\
        \hline
        8 & Process, Plant, Machine Operatives & 188   \\
        \hline
        9 & Elementary Occ.& 124   \\
 \hline
    \end{tabularx}
    \caption{The table shows the major groups (left column) and classified jobs with multiple sub-major
groups (middle column) by Standard Occupation Classification. The right-most column represents the number of main users \cite{pan2019twitter}.}
\label{table:4}
\end{table}

In order to construct the social network, each follower/following relationship is considered as an undirected edge for predicting the occupational classes of the main users \cite{pan2019twitter}. In this social network, the main Twitter users are considered as being connected with one another via common followers and following (follow) IDs. The follow IDs which only connect a few of the main IDs are considered to be weak since the flow of information between the main user IDs will be less than these follow IDs. As a result, the network was further refined by keeping only the follow IDs which have more than $10$ connections to the main user IDs. After performing the refining step an unweighted graph was constructed in which all the main IDs were connected to each other and the refined graphs consisted of $34,630$ unique users including the 4557 main users. In the network, only $2550$ main user IDs have at least one direct connection with another main user ID. Thus, when constructing the network model the main user IDs often shared common follow IDs which enabled the researchers to extract rich network information.

\subsection{Non-Twitter Based Datasets}
Datasets for validating homophilous models were generated from other types of social media platforms as well such as the Weibo Sina website which is a Chinese microblogging site\cite{yang2015birds}. Similarly, Furthermore, synthetic networks were also developed by using an artificial undirected network. However, the artificial dataset was not thoroughly described in depth \cite{karimi2018homophily}. Datasets: $3$, $7$, and $8$ from Table \ref{table:33}, shows the datasets generated from other sources. The following paragraphs give an overview of how these datasets were developed for analyzing the effects of homophily.

\subsubsection{Asian Telecom Dataset}
For studying the effects of homophily for product purchasing \cite{ma2015latent}, purchases of Caller Ring-Back Tones (CRBT) data were provided by an Asian mobile network. This network data was mainly used for predicting consumers’ product choice decision and purchase timings. The dataset consisted of three months of detailed $300$ million phone call histories of the company’s approximately $3.7$ million customers. The call attributes were the caller or callee phone numbers and the duration of the phone conversation. The CRBT product was bought by 750,000 customers. The pattern of the CRBT purchasing data was explored to find out the main driving forces of the customers to buy the product. The communication between friends were tracked and analyzed using this dataset. When friends of the customer get exposed to the ringtone by calling them and purchase the same product. Moreover, CRBT is a cheap and economical product that has been purchased by more than $750,000$ customers so the researchers claimed that the dataset of phone call histories provides a convenient platform for studying communications on this product.

\subsubsection{ Weibo Users Dataset}
In order to validate the proposed topic model for link prediction \cite{yang2015birds}, Data were extracted from the Sina Weibo\footnote{\href{https://www.weibo.com}{https://www.weibo.com}}. The dataset contains about $2000$ verified users, in which each user is represented by a single document. The link information between the pairs of users was also collected, when both the users’ posts were present in the dataset. The link information refers to three types of interactions which include mentioning, retweeting, and following in the Weibo website.

Generally, Twitter Search API is used to generate the desired network data for validating the proposed homophilous models. Users and their textual tweets are mainly used as features to develop the dataset. Social media platforms have other features such as images and videos which are posted by users. However, such features have not yet been included in the datasets for evaluating homophily. Furthermore, the size of the datasets varies a lot, ranging from $2000$ users’ posts to $75,000$ users’ posts. Yet, no benchmark has been set to have a minimum standard size of dataset to validate any of the proposed models. Other than using Twitter Search API, microblogs, and artificial data such as Weibo Sina and artificial undirected network has also been used to generate the network data.

\section{Conclusion}\label{Conclusion}
The homophily principle in the domain of social network analysis is an important concept that has been studied broadly. It has been used to examine the behavior of users on social media platforms. Generally, in social media, users tend to connect with others where they have similar interests with one another. Several studies have been carried out to study the homophilic patterns especially among majority and minority groups. These studies have deduced that majority and larger groups receive information faster than smaller, minority groups. The information reaches like-minded individuals quicker. Besides, multiple types of models have been proposed for measuring the degree of homophily. Network modeling and topic modeling have been mainly used for analyzing the strength of a relationship of a user with the user's nearest neighbors. Network modeling was conducted on network features and topic modeling was conducted on users' textual tweets respectively. Recently, the effect of homophily has also been studied by combining textual and network features with a highly efficient neural network model. However, the content of the interactions occurring between users in social media are still inadequately understood\cite{Hywel}. Furthermore, the content of social media ranges from texts to videos, which needs different types of analysis. Most of the studies focused either on textual posts, hashtags tweeted by users, mentions of users, or users' network connections. However, there is a research gap as comprehensive studies have not been conducted, concerning images and videos posted by users. Whether these contents of social media have any effect on the degree of homophily in online platforms is not fully understood as of yet. Therefore, state-of-art methods should also focus on measuring the level of homophily by using these features.

In this paper, we presented a survey on the usage of the Homophily principle for computing social network analysis. This thorough survey will enable researchers to explore new methods to measure the degree of homophily. In summary, multiple methodologies have been proposed over the years to measure the level of homophily by using different types of modeling approaches. This includes topic modeling, network modeling, ERGMs, where each of the models has its own merits and drawbacks. These models are validated by either using synthetic data or by using real-life data from social networking sites such as Twitter. However, the range of data used by each of the proposed models varied to a great extent ranging from posts from only 2000 users to 75,000 users' posts. Therefore, for better comparability of different features and methods, we argue for a benchmark dataset for homophily detection.

\section*{Acknowledgments} 
We would like to thank the reviewers for their
helpful comments on our work. This work is supported by Natural Sciences and Engineering Research Council of Canada (NSERC).
\bibliographystyle{IEEEtran}
\bibliography{sample-base}
\onecolumn

\newpage
\newpage
\appendix 

\section{Table of References}\label{appendix:a}
\begin{center}
  Table 3: The information of all the references selected for the survey  
\end{center}
\begin{small}
\begin{center}
\begin{longtable}{|p{5.5cm}|p{5.5cm}|p{1.5cm}|p{1.25cm}|p{0.75cm}|p{0.75cm}|}

\hline
\rowcolor[HTML]{343434}
\color[HTML]{FFFFFF}
         Title & \color[HTML]{FFFFFF}Venue & \color[HTML]{FFFFFF}Citations &\color[HTML]{FFFFFF} Quartile & \color[HTML]{FFFFFF} H-index &\color[HTML]{FFFFFF}Year \\\hline
\endfirsthead
\multicolumn{4}{c}%
{\tablename\ \thetable\ -- \textit{Continued from previous page}} \\
\hline
\rowcolor[HTML]{343434}
\color[HTML]{FFFFFF}
         Title & \color[HTML]{FFFFFF}Venue & \color[HTML]{FFFFFF}Citations &\color[HTML]{FFFFFF} Quartile &  \color[HTML]{FFFFFF} H-index & \color[HTML]{FFFFFF}Year \\\hline
\hline
\endhead
\hline \multicolumn{4}{r}{\textit{Continued on next page}} \\
\endfoot
\hline
\endlastfoot
         Friendship as a social process: A  substantive and methodological  analysis \cite{lazarsfeld1954friendship} & Freedom and control in modern society  & 3069 & - & - & 1954\\
        \hline
         Birds of a feather: Homophily in social networks \cite{mcpherson2001birds} & Annual review of sociology & 15216 & - & 151 & 2001\\
        \hline
         Why does everybody hate me?  balance,status, and homophily: The triumvirate of signed tie formation \cite{yap2015does} & Social Networks & 43 & Q1 & 85 & 2015\\
        \hline
         How social ties transcend class boundaries ? Network variability as tool for exploring occupational homophily \cite{cepic2020social} &Social Networks& - & Q1  & 85 & 2020\\
        \hline 
         Quantifying segregation in an integrated urban physical-social space \cite{xu2019quantifying} &Journal of the Royal Society Interface  & - & Q1 & 114 & 2019\\
        \hline
         Trustworthy health-related tweets on social media in Saudi Arabia: tweet metadata analysis \cite{albalawi2019trustworthy} &Journal of medical Internet research & 1 & Q1 & 116 & 2019\\
        \hline
         For better or for worse? A systematic review of the evidence  on social media use and depression among lesbian, gay, and bisexual minorities \cite{escobar2018better} &Journal of medical Internet research  & 6 & Q1 & 116 & 2019\\
        \hline
         Social media and human need satisfaction: Implications for social media marketing \cite{zhu2015social} & Business Horizons & 194 & Q1 & 67 & 2015\\
        \hline
         Will they come and will they stay? Online social networks and news consumption on external websites \cite{mahmood2017will} & Business Horizons & 194 & Q1 & 67 & 2015 \\
         \hline
         Political homophily in social relationships: Evidence from online dating behavior \cite{huber2017political} & The Journal of Politics & 100 & -& 50 & 2017 \\
         \hline
         Homophily of music listening in online social networks of China \cite{zhou2018homophily} & Social Networks & 5 & Q1 &85 & 2018\\
         \hline
         Latent homophily or social influence? An empirical analysis of purchase within a social network \cite{ma2015latent} & Management Science & 85 & Q1 & 221 &2016\\
         \hline
         Twitter Homophily: Network Based Prediction of User’s Occupation \cite{pan2019twitter} & Proceedings of the 57th Annual Meeting of the Association for Computational Linguistics  & - & - & 51  &2019\\
         \hline
         Homophily, group size, and the diffusion of political information in social networks: Evidence from Twitter \cite{halberstam2016homophily} & Journal of public economics & 178 & Q1 & 124 & 2016\\
         \hline
         Analyze users' online shopping behavior using interconnected online interest-product network \cite{han2018analyze} & WCNC & 1 &- & 80  &2018\\
         \hline
         Race, school integration, and friendship segregation in America \cite{moody2001race} & American journal of Sociology & 1330 & Q1 & 160 & 2001\\
         \hline
         Shared contexts, shared background, shared values--Homophily in Finnish parliament members’ social networks on Twitter \cite{koiranen2019shared} & Telematics \& Informatics & 3 & Q1 & 52 & 2019\\
         \hline
         Building the community: Endogenous network formation, homophily and pro social sorting among therapeutic community residents \cite{warren2020building} & Drug and Alcohol Dependence & - & Q1 &151 & 2020\\
         \hline
         Purity homophily in social networks \cite{dehghani2016purity} & Journal of Experimental Psychology: General & 83 & Q1 & 138  &2016\\
         \hline
         Structural transition in social networks: The role of homophily \cite{murase2019structural} & Scientific reports & 1 & Q1 & 149 & 2019\\
         \hline
         Information diffusion and opinion change during the gezi park protests: Homophily or social influence? \cite{dincelli2016information} & Database: The Journal of biographical logical Databases and Curation & 88& - & 65 &2016\\
        \hline
        Trip distribution modeling with Twitter data \cite{pourebrahim2019trip} & Computers, Environment and Urban Systems  & 2 & Q1 & 74 & 2019\\
        \hline
         Good Games, bad host? Using big data to measure public attention and imagery of the Olympic Games \cite{kassens2019good} & Cities & 5 & Q1 & 74 & 2019\\
        \hline
        Using Facebook for Qualitative Research: A Brief Primer \cite{franz2019using} & Journal of medical Internet research & - & Q1 & 116 & 2019\\
         \hline
        Sensitivity analysis for contagion effects in social networks \cite{VanderWeele} & Sociological Methods \& Researchs & 124 & Q1 & 65 & 2011\\
        \hline
         Birds of a schedule flock together: Social networks, peer influence, and digital activity cycles \cite{liang2018birds} & Computers in Human Behavior & 3  & Q1 & 137 & 2018\\
         \hline
         Social media engagement: What motivates user participation and consumption on YouTube? \cite{khan2017social} & Computers in Human Behavior & 254 & Q1 &137& 2017\\
         \hline
      Social media and web presence for patients and professionals: evolving trends and implications for practice \cite{barreto2017social} & PM\&R & 31 & -&53 &2017\\
         \hline
        The media inequality: Comparing the initial human-human and human-AI social interactions \cite{mou2017media} & Computers in Human Behavior & 59 & Q1 &137 & 2017\\
         \hline
         Sentiment analysis on Twitter: A text mining approach to the Syrian refugee crisis \cite{Nazan} & Telematics and Informatics & 80 & Q1 & 52 &2018\\
         \hline
        Network analysis reveals open forums and echo chambers in social media discussions of climate change \cite{Hywel} & Global environmental change & 201 & Q1 & 147 &2015\\
         \hline
         The ambivalence of cultural homophily: Field positions, semantic similarities, and social network ties in creative collectives \cite{basov2019ambivalence} & Poetics	 & 4 &Q1 &54 &2019\\
         \hline
         A survey on deep learning in medical image analysis & Medical image analysis \cite{litjens2017survey} & 2991 & Q1 &113 &2017\\
         \hline
         Deep learning in agriculture: A survey \cite{kamilaris2018deep} & Computers and electronics in agriculture & 434 & Q1 & 96 &2018\\
         \hline
         Graph embedding techniques, applications, and performance: A survey \cite{goyal2018graph} & Knowledge-Based Systems & 521 & Q1& 94 &  2018\\
         \hline
         Deep learning based recommender system: A survey and new perspectives \cite{zhang2019deep} & ACM Computing Surveys & 437 & Q1 & 132 & 2019\\
          \hline
         Homophily, structure, and content augmented network representation learning \cite{zhang2016homophily} & 2016 IEEE 16th international conference on data mining (ICDM) & 55 & Q1 & 100 & 2016\\
          \hline
         The social structure of political echo chambers: Variation in ideological homophily in online networks \cite{boutyline2017social} & Political Psychology & 150 & Q1 & 80 & 2019\\
          \hline
         Equality of opportunity in classification: A causal approach \cite{zhang2018equality} & Advances in Neural Information Processing Systems & 13 & - & 54 & 2018\\
          \hline
         Yes, there is a correlation: -from social networks to personal behavior on the web \cite{singla2008yes} & Proceedings of the 17th international conference on World Wide Web & 344 & - & 64 & 2008\\
          \hline
         The old boy (and girl) network: Social network formation on university campuses \cite{mayer2008old} & Journal of public economics & 443 & Q1 & 123 & 2008\\
          \hline
         YouTube vloggers’ popularity and influence: The roles of homophily, emotional attachment, and expertise \cite{ladhari2020youtube} & Journal of Retailing and Consumer Services & - & Q1 & 65 & 2020\\
          \hline
         Dynamic social balance and convergent appraisals via homophily and influence mechanisms \cite{mei2019dynamic} & Automatica & 1 & Q1 & 239 & 2019\\
          \hline
         Dominant frames in legacy and social media coverage of the IPCC Fifth Assessment Report \cite{o2015dominant} & Nature Climate Change & 151 & Q1 & 136 & 2015\\
          \hline
         Twitter users change word usage according to conversation-partner social identity \cite{tamburrini2015twitter} & Social Networkss & 47 & Q1 & 85 & 2015\\
          \hline
         Stance polarity in political debates: A diachronic perspective of network homophily and conversations on Twitter \cite{lai2019stance} & Data \& Knowledge Engineering & 1 & Q2 & 79 & 2019\\
          \hline
         Valence-based homophily on Twitter: Network analysis of emotions and political talk in the 2012 presidential election \cite{himelboim2016valence} & New media \& society & 63 & Q1 & 87 & 2016\\
          \hline
         Hashtag homophily in twitter network: Examining a controversial cause-related marketing campaign \cite{xu2020hashtag} & Computers in Human Behavior & - & Q1 & 137 & 2020\\
          \hline
         Polarized frames on “climate change” and “global warming” across countries and states: Evidence from Twitter big data \cite{jang2015polarized} & Global Environmental Change & 132 & Q1& 147 & 2015\\
          \hline
         Homophily influences ranking of minorities in social networks \cite{karimi2018homophily} & Scientific reports & 20 & Q1 & 149 & 2018\\
          \hline
         Status seeking and perceived similarity: a consideration of homophily in the social servicescape \cite{hanks2017status} &International Journal of Hospitality Management& 29 & Q1 & 93 & 2017\\
          \hline
         Emergence of scaling in random networks \cite{barabasi1999emergence} & Science & 36019 & Q1 & 1058 & 1999\\
          \hline
         Spectral centrality measures in complex networks \cite{perra2008spectral} & Physical Review E & 148 & Q1 & 190 & 2008\\
          \hline
         The influence of cause-related marketing on consumer choice: does one good turn deserve another? \cite{barone2000influence} &Journal of the academy of marketing Science & 1383 & Q1 & 148 & 2000\\
          \hline
         Latent semantic indexing: A probabilistic analysis \cite{papadimitriou2000latent} & Journal of Computer and System Sciences & 1280 & Q2 & 81 & 2000\\
          \hline
         Latent dirichlet allocation \cite{blei2003latent} & Journal of machine Learning research & 31189 & Q1 & 173 & 2003\\
          \hline
         Exponential random graph model parameter estimation for very large directed networks \cite{stivala2020exponential} &PloS one & 4 & Q1 & 268 & 2020\\
          \hline
         Measuring the impact of spammers on e-mail and Twitter networks \cite{colladon2019measuring} & International Journal of Information Management & 10 & Q1 & 91 & 2019\\
          \hline
         Friend or frenemy? Experiential homophily and educational track attrition among premedical students \cite{grace2018friend} & Social Science \& Medicine & 1 & Q1 & 213 & 2018\\
          \hline
         A belief-based theory of homophily \cite{kets2019belief} & Games and Economic Behaviory & 5 & Q1 & 84 & 2019\\
          \hline
         Tweeting for social justice in\# Ferguson: Affective discourse in Twitter hashtags \cite{blevins2019tweeting} & new media \& society & 3 & Q1 & 87 & 2019\\
          \hline
         An adaptive temporal-causal network model for social networks based on the homophily and more-becomes-more principle \cite{van2019adaptive} & Neurocomputing & 7 & Q1 & 110 & 2019\\
          \hline
         Semi-supervised classification with graph convolutional networks \cite{kipf2016semi} & arXiv & 3196 & - & - & 2016\\
          \hline
         An analysis of the user occupational class through Twitter content \cite{preoctiuc2015analysis} & Proceedings of the 53rd Annual Meeting of the Association for Computational Linguistics and the 7th International Joint Conference on Natural Language Processing (Volume 1: Long Papers) & 134 & - & 51 & 2015\\
          \hline
         A simple model of homophily in social networks \cite{currarini2016simple} & European Economic Review & 66 & Q1 & 116 & 2016\\
          \hline
         Birds of a feather linked together: A discriminative topic model using link-based priors \cite{yang2015birds} & Proceedings of the 2015 Conference on Empirical Methods in Natural Language Processing & 5 & - & 88 & 2015\\
          \hline
         Effect of homophily on network formations \cite{kim2017effect} & Communications in Nonlinear Science and Numerical Simulation & 21 & Q1 & 96 & 2017\\
          \hline
         MedLDA: maximum margin supervised topic models \cite{zhu2012medlda} & Journal of Machine Learning Research & 443 & Q1 & 173 & 2012\\
          \hline
         Topic models conditioned on arbitrary features with dirichlet-multinomial regression \cite{mimno2012topic}& arXiv & 389 & - & - & 2012\\
          \hline
        Gibbs max-margin topic models with data augmentation \cite{zhu2014gibbs} & The Journal of Machine Learning Research & 75 & Q1 & 173 & 2014\\
          \hline
         Lexical and hierarchical topic regression \cite{nguyen2013lexical} & Advances in neural information processing systems & 61 & - & 54 & 2013\\
          \hline
         Distributed representations of words and phrases and their compositionality \cite{mikolov2013distributed} & Advances in neural information processing systems & 18726 & - & 54 & 2013\\
           \hline
          A density-based method for adaptive LDA model selection \cite{cao2009density} & Neurocomputing & 276 & Q1 & 110 & 2009\\
           \hline
         Reading tea leaves: How humans interpret topic models \cite{chang2009reading} & Advances in neural information processing systems& 1668 & - & 54 & 2009\\
           \hline
        Optimizing semantic coherence in topic models \cite{mimno2011optimizing} & Proceedings of the conference on empirical methods in natural language processing & 918 & - & 88 & 2011\\
           \hline
         The effect of calorie posting regulation on consumer opinion: A flexible latent Dirichlet allocation model with informative priors \cite{puranam2017effect} & Marketing Science & 34 & Q1 & 113 & 2017\\
           \hline
         Specification of exponential-family random graph models: terms and computational aspects \cite{morris2008specification} & Journal of statistical software & 302 & Q1 & 115 & 2008\\
           \hline
         Exponential random graph models for social networks: Theory, methods, and applications \cite{lusher2013exponential} & Cambridge University Press & 705 & - & - & 2013\\
           \hline
         Opening the black box of link formation: Social factors underlying the structure of the web \cite{gonzalez2009opening} & Social Networks & 77 & Q1 & 85 & 2009\\
           \hline
         An introduction to exponential random graph (p*) models for social networks \cite{robins2007introduction} & Social Networks & 1677 & Q1 & 85 & 2007\\
           \hline
        Reconsidering power in multi stakeholder relationship management \cite{saffer2018reconsidering} & Management Communication Quarterly & 12 & Q1 & 55 & 2018\\
           \hline
        Why the Bass model fits without decision variables \cite{bass1994bass} & Marketing Science & 1044 & Q1 & 113 & 1994\\
           \hline
         Tweetmotif: Exploratory search and  topic summarization  for  twitter \cite{o2010tweetmotif} & Fourth  International  AAAI Conference on Weblogs and Social Media, 2010 & 411 & - & 60 & 2010\\
           \hline
        A method of establishing groups of equal amplitude in plant sociology based on similarity of species content and its application to analyses of the vegetation on Danish commons \cite{sorensen1948method} & Journal of Machine Learning Research & 2871 & - & - & 1948\\
          
\label{table:44}
\end{longtable}
\end{center}
\end{small}

\end{document}